\documentclass[11pt,twoside]{article}
\usepackage{./asp2014}
\usepackage{color}\usepackage{graphicx}
\usepackage[normalem]{ulem}

\aspSuppressVolSlug
\resetcounters

\begin{document}

\title{The Molecular High-z Universe on Large Scales: Low-surface-brightness CO and the strength of the ngVLA Core}
\author{Bjorn Emonts$^1$, Chris Carilli$^2$, Desika Narayanan$^{3,4,5}$, Matthew Lehnert$^6$, Kristina Nyland$^1$}
\affil{$^1$National Radio Astronomy Observatory, 520 Edgemont Road, Charlottesville, VA 22903 \email{(bemonts@nrao.edu)}}
\affil{$^2$National Radio Astronomy Observatory, P.O. Box 0, Socorro, NM 87 801}
\affil{$^3$Department of Astronomy, University of Florida, 211 Bryant Space Science Center, Gainesville, FL 32611}
\affil{$^4$University of Florida Informatics Institute, 432 Newell Drive, CISE Bldg E251, Gainesville, FL 32611}
\affil{$^5$Cosmic Dawn Centre (DAWN), University of Copenhagen, Julian Maries vej 30, DK-2100, Copenhagen, Denmark}
\affil{$^6$Sorbonne Universit\'{e}, CNRS, UMR 7095, Institut d'Astrophysique de Paris, 98bis bvd Arago, 75014, Paris, France}

\begin{abstract}
The Next-Generation Very Large Array (ngVLA) will revolutionize our understanding of the Early Universe by tracing the coldest phase of molecular gas -the raw ingredient for star formation- in the most distant galaxies and galaxy-clusters. The km-scale core of the ngVLA will be densely packed with antennas, making it a prime instrument for imaging low-surface-brightness emission from large-scale molecular gas in the high-$z$ circum- and inter-galactic medium (CGM/IGM). Recent studies indicate that large amounts of cold molecular gas are hiding in the 10s$-$100 kpc environments of distant galaxies, but that technical limitations on existing telescope arrays have prevented us from efficiently detecting these large molecular reservoirs. This may have led to a severely biased view of the molecular Universe. We present the science case for low-surface-brightness CO observations of the Early Universe, and how the core of the ngVLA will reveal the cold molecular Universe to limits and at scales not currently detectable by radio telescopes. As such, the ngVLA core will be a powerful instrument for studying the cold baryon cycle that drives the early evolution of galaxies and clusters.
\end{abstract}

\section{Introduction}

The evolution of galaxies is tightly linked to processes in the circum- and inter-galactic medium (CGM/IGM). Unfortunately, most of the baryons in the CGM/IGM are too faint to be easily detected. At high-$z$, we view glimpses of dark baryonic halos through quasar absorption lines, or cooling-radiation emitted as Ly$\alpha$. Absorption-line work inferred the presence of large, $\sim$100 kpc halos of warm/cool (T$\sim$10$^{4}$\,K), metal-enriched gas around high-$z$ quasars \citep[e.g.,][]{pro14,lau16}. However, a direct connection to the stellar growth of massive galaxies remains missing, because we have yet to identify the ultimate reservoir of halo gas that has sufficient mass to fuel widespread star-formation, namely the cold molecular gas (T$\sim$10$-$100 K).

The massive Spiderweb Galaxy at $z$=2.2 \citep{mil06} revealed the first evidence for the existence of a cold molecular CGM in the distant Universe \citep[][Fig.\,1]{emo16}. CO(1-0) observations with very compact array configuration of the Australia Telescope Compact Array (ATCA) traced a 70 kpc reservoir of molecular halo gas across the CGM of this massive forming proto-cluster galaxy \citep{emo16}. The extended CO follows diffuse blue light from star formation that occurs in-situ within the CGM \citep[][see Fig.\,1]{hat08}. This is evidence that there is a massive (M$_{\rm H2}$\,$\sim$\,10$^{11}$\,M$_{\odot}$) reservoir of gas in the CGM that cooled well beyond the temperature of Ly$\alpha$-emitting gas (T $\sim$ few 10$^{3}$ $-$ 10$^4$ K), and is actively feeding star formation across the halo. More recent observations of [C\,I] $^{3}$P$_{1}$-$^{3}$P$_{0}$ and CO(4-3) with the Atacama Large Millimeter Array (ALMA) revealed that the cold CGM has densities of several 100 cm$^{-3}$, with average carbon abundance and excitation conditions resembling those of starforming galaxies \citep{emo18}. This does not corroborate models of efficient and direct stream-fed accretion of relatively pristine gas \citep{dek09}. Instead, the [CI] and CO properties agree with more complex recycling models, where the gas in the IGM is a melange from various sources $-$ metal-enriched outflows, mass transfer among galaxies, gas accretion, and mergers \citep[see][]{emo18}. 

Other studies using very compact interferometers also found CO-emitting gas on scales of tens of kpc around high-$z$ massive galaxies \citep[e.g.,][]{emo14,cic15,emo15,gin17,dan17}. These results are starting to reveal a picture that the CGM is truly multi-phase, and that the Early Universe may contain much more molecular gas hiding outside galaxies than has thus far been observed. However, many questions remain about the nature of the cold molecular CGM/IGM, and its role in early galaxy evolution. How common is the presence of a cold CGM, in particular among the more general population of high-$z$ galaxies? Is the cold gas diffuse or associated with star-formation? What is its origin (tidal debris, outflows, mixing, or accretion)? And how can we observe this cold CGM/IGM? 

In this paper, we will explain that technical limitations on array designs have mostly prevented us from reaching the sensitive low-surface-brightness levels needed to efficiently image widespread molecular gas across the CGM/IGM. This likely produced a severely biased view of the molecular Universe. The core of the Next-Generation VLA (ngVLA) will be a powerful instrument for tracing the molecular CGM, allowing us to study the cold baryon cycle that governs the early growth of massive galaxies.

\articlefigure[width=0.76\textwidth]{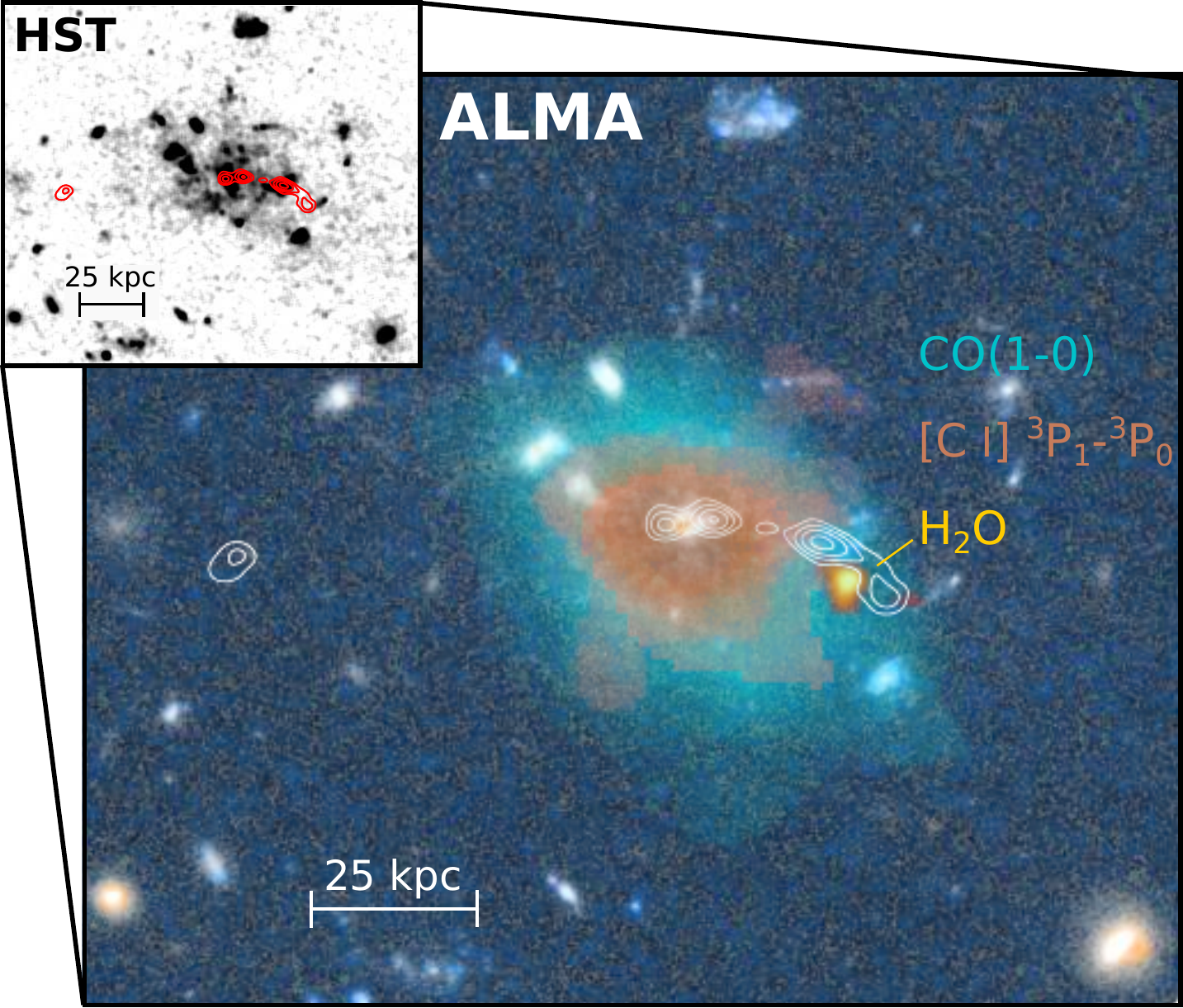}{fig:spiderweb}
{The cold circum-galactic medium of the Spiderweb Galaxy ($z$\,=\,2.2). A total intensity image of the CO(1-0) emission is shown in light-blue in the main panel, as detected with the ATCA \citep{emo16}. Superposed in brown are various channel maps of atomic carbon, [C\,{\small I}] $^3$P$_{1}$\,$-$\,$^3$P$_{0}$, as detected with ALMA \citep{emo18}. A bright spot of H$_{2}$O emission is shown in yellow \citep{gul16}. The background image is taken with the {\it Hubble Space Telescope} Advanced Camera for Surveys ({\it HST}/ACS) in the F475W and F814W filters \citep{mil06}. The radio source is shown in white contours, as observed with the VLA in the C-configuration at 35 GHz \citep{emo16}. The in-set in the top-left corner shows the combined {\it HST}/ACS F475W+F814W image of the same region as the main panel \citep{mil06}, but optimized to visualize the diffuse optical light from in-situ star formation \citep{hat08}.}

\section{Low-surface-brightness CO}

Some of the strongest tracers for molecular gas are the various rotational transitions of carbon-monoxide, $^{12}$CO($J$,$J$-1). With the Atacama Large Millimeter Array (ALMA), it has become routine to observe CO, and other molecular species, in high-$z$ galaxies. However, imaging the cold molecular medium outside galaxies on tens to hundred kpc scales has been limited by two observational challenges:\\
\ \\
$\bullet$ {\bf Excitation conditions of the molecular medium.} The critical density of CO($J$,$J$-1) scales roughly with n$_{\rm crit}$\,$\propto$\,$J^{3}$. The ground-transition CO(1-0) ($\nu_{\rm rest} = 115.27$ GHz) has an effective critical density of only several 100 cm$^{-3}$, and the $J$\,=\,1 level of CO is substantially populated down to $T$\,$\sim$10 K \citep{pap04}. However, these values increase by an order of magnitude or more for the high-$J$ transitions, like CO(3-2) and higher. This means that these high-$J$ transitions trace the warmer and denser molecular gas in starburst and AGN regions, and may severely underestimate masses from colder, lower density, and sub-thermally excited gas components \citep{pap00}. To get robust estimates of the {\sl total} molecular gas mass in galaxies at $z>$1.5, it is vital that CO(1-0) or CO(2-1) are observed in the 20-50 GHz regime. {\sl This is particularly important for widespread CO-emitting reservoirs of molecular gas in the circum- and inter-galactic medium, as the excitation conditions and temperature of the molecular gas in these environments may be low compared to the ISM of galaxies.}

\articlefigure[width=0.97\textwidth]{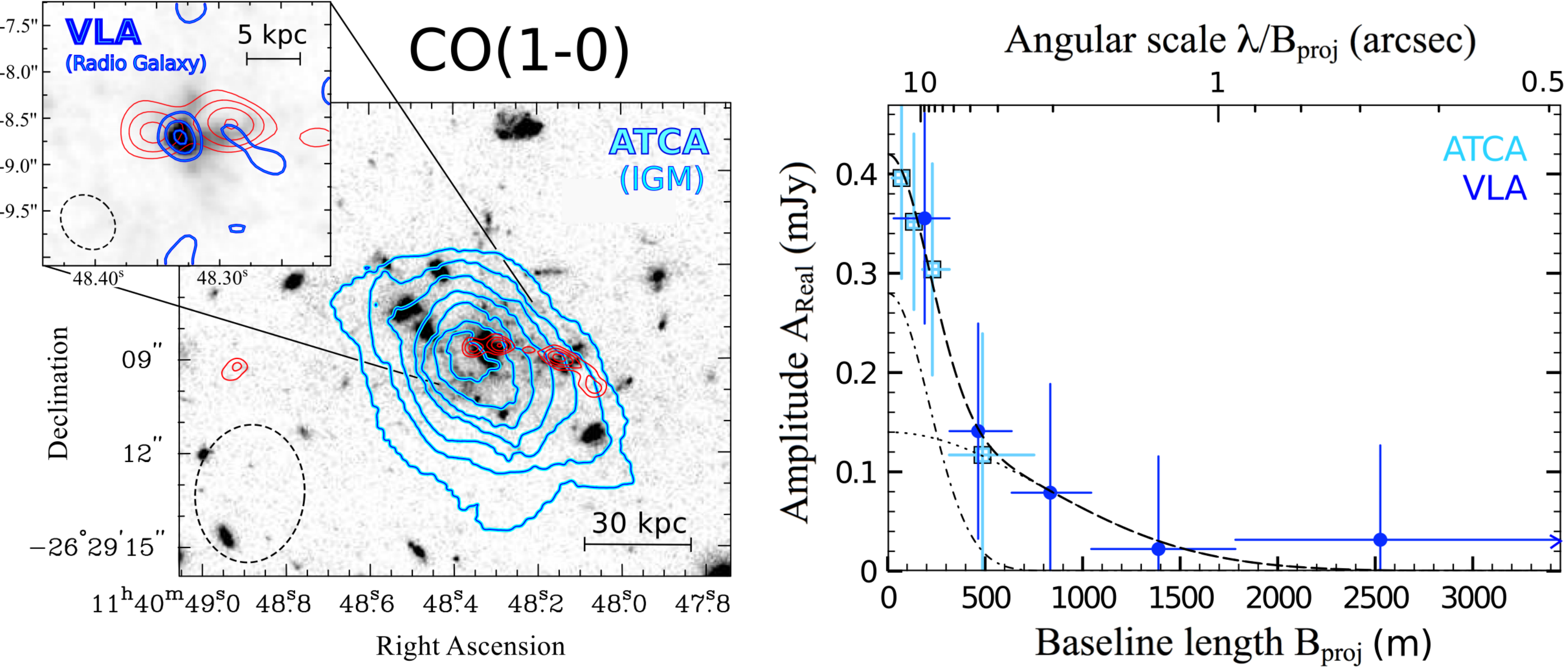}{fig:spiderweb2}
{Low-surface-brightness CO(1-0) from Fig.\,1 across the halo of the Spiderweb Galaxy ($z$=2.2), reproduced from \citet{emo16} Left: ATCA CO(1-0) contours (blue) are overlaid onto the {\it HST}/ACS F475W+F814W image from \citet{mil06}. Contour levels: 0.020, 0.038, 0.056, 0.074, 0.092, 0.110, 0.128 Jy\,beam$^{-1}$ $\times$ km\,s$^{-1}$. The inset shows the CO(1-0) total-intensity contours from the VLA in DnC+CnB-configuration at 3.5, 4.5$\sigma$. The full-resolution VLA data reveal only a third of the total CO imaged on large scales with ATCA. Right: Visibility-amplitudes of the CO(1-0) emission are plotted as function of the projected baseline length in data from the ATCA (light-blue squares) and VLA DnC+CnB (dark-blue dots). As described in \citet{emo16}: "The amplitude plotted on the Y-axis is the real part of the complex interferometer visibility for the CO-signal, when vector-averaged across the velocity range -100 to +200 km/s and over all baselines covered by the horizontal error bars. The $\pm$1$\sigma$ uncertainty in the amplitude of the averaged signal is indicated with the vertical error bars. The dashed line shows a model where we superimposed a Gaussian distribution with FWHM = 3'' (dash-dotted line) and a Gaussian distribution with FWHM = 0.8'' (dotted line)." This plot illustrates that it is essential to use short-baseline configurations to recover extended CO emission. }
\ \\
\noindent $\bullet$ {\bf Short-spacing problem of radio interferometers.} Radio interferometers sample the sky at discrete intervals, set by the length of antennas pairs (`baselines'), frequency coverage and scan-time. Because large-scale features are filtered out by widely separated antennas, {\sl low-surface-brightness emission can only be imaged with compact array-configurations at relatively low spatial resolution.} This problem is particularly severe when studying extended gas reservoirs in the mm regime. For example, to recover CO at $z\sim$2 on scales $\ge$15 kpc requires antennas to be placed $<$1\,km apart at 35 GHz, and less than a few 100m for the lower ALMA bands! This problem is visualized in Fig. 2.\\
\ \\
\noindent Unfortunately, millimeter observatories, like ALMA, generally operate at frequencies $>$85 GHz. This means that the ground-transition CO(1-0) rapidly redshifts out of the receiver bands. Several of the traditional cm-observatories, like the Very Large Array (VLA) and ATCA, can target redshifted CO(1-0) and CO(2-1) in the 20-50 GHz regime, but they crucially lack sensitivity on very short ($<<$1\,km) baselines for efficiently imaging the cold molecular CGM/IGM (Fig.\,2).

\articlefigure[width=\textwidth]{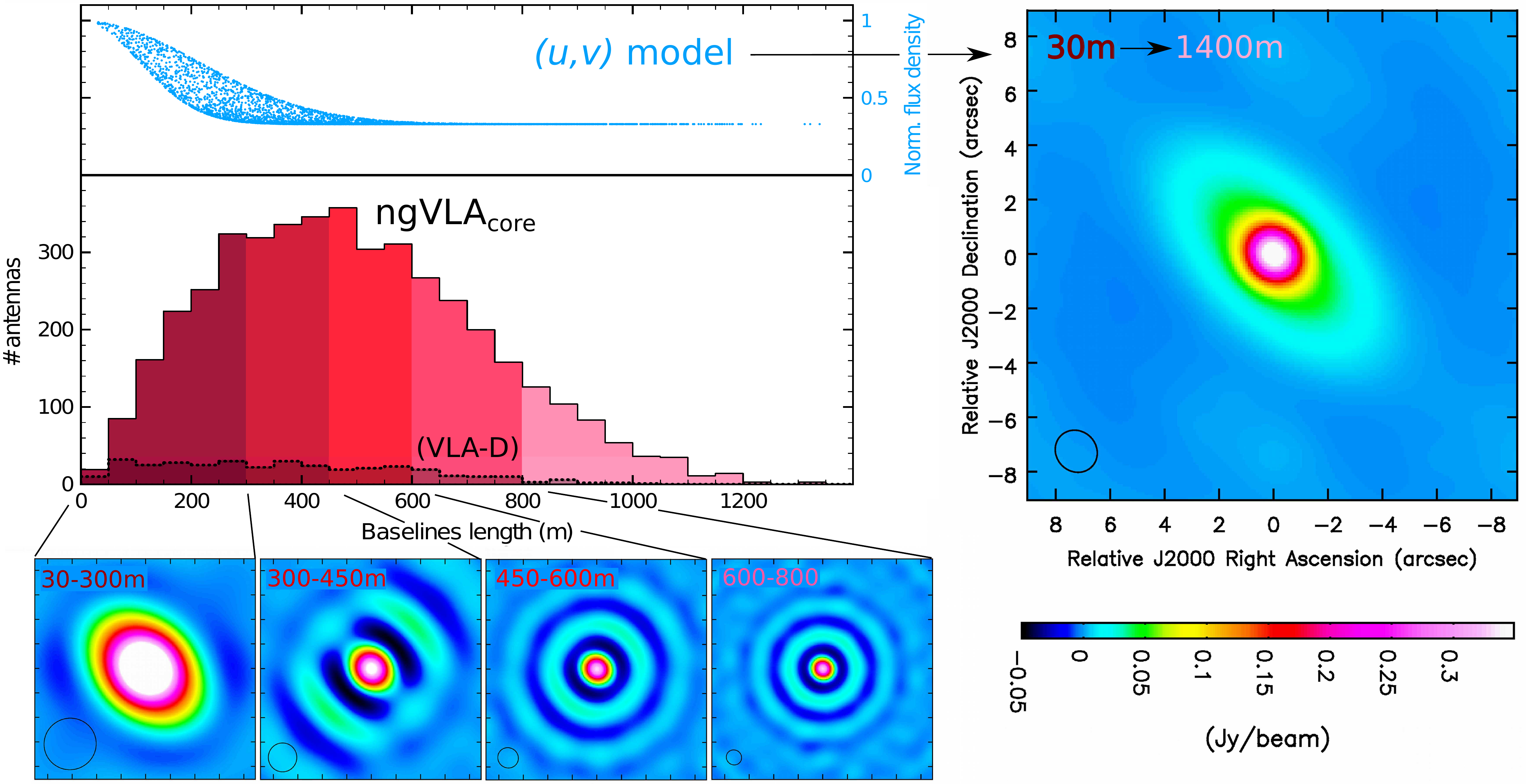}{fig:baselines}
{Simulated observations with the ngVLA core. {\sl Top left:} Red-colored histogram of the baseline distribution of the 18m antennas in the ngVLA core, using the reference design of \citet{car18}. The colors from dark to light red contain 24$\%$ (30$-$300m), 23$\%$ (300$-$450m), 22$\%$ (450$-$600m), 20$\%$ (600$-$800m) and 11$\%$ (800$-$1400m) of the total number of baselines in the core. The overlaying black histogram at the bottom shows the baseline distribution of the VLA in D-configuration. The blue dots at the top of the figure are drawn on the same x-axis and represent a simulated snap-shot observation of a target model at the zenith, using the simulation tools in the Common Astronomy Software Applications (CASA) software \citep{mcm07}. This model consists of a Gaussian function that represents extended emission with a superposed delta-function of an unresolved point-source component, both placed at the phase center. The Gaussian has FWHM $b_{\rm maj}$\,=\,6$^{\prime\prime}$ and $b_{\rm min}$\,=\,3$^{\prime\prime}$ with PA\,=\,45$^{\circ}$, which corresponds to about 50\,$\times$\,25 kpc at $z$\,=\,2. The signal is normalized and placed in a single channel of width 30 MHz, with 2/3rd (1/3rd) of the total flux in the Gaussian (point-source) component. No noise has been added to the simulation and a basic clean of 1000 iterations down to 0.1 $\times$ the peak flux was applied. {\sl Top right:} Total intensity image of the simulated emission of the target model using the ngVLA core with an exposure time of 8h. The image includes all the baselines in the range 30$-$1400m. {\sl Bottom:} The small panels show the total intensity image when restricting the uv-coverage to the baselines indicated in the colored parts of the histogram (i.e., 30$-$300m, 300$-$450m, 450$-$600m and 600$-$800m). The dimensions and color-scaling for all four panels are the same as for the main image of the target model in the top right panel. This figure illustrates that the shortest ($\le$ several 100m) baselines in the core of the ngVLA are crucial for recovering and imaging extended low-surface-brightness emission.}

\section{The strength of the ngVLA core}

The core of the next-generation VLA (ngVLA) \citep{car16} will be densely packed with antennas that will vastly supersede the current VLA D-configuration in number of short baselines. The current reference design for the core includes 94 antennas with a diameter of 18m that cover baselines out to about 1 km \citep{car18}. While the longer ngVLA baselines will not be able to detect extended CO because they resolved out the emission, the ngVLA core will be a revolutionary new instrument for unlocking the cold molecular Universe.  This is further visualized in Fig.\,3, where we show the baseline coverage of the ngVLA core \citep{car18}, as well as simulated imaging of extended emission. The simulation shown in Fig.\,3 is simplistic, and does not take into account noise or gas kinematics and clumpiness. Nevertheless, it shows the need for sufficient sensitivity on baselines less than several $\times$ 100m in the core of the ngVLA, in order to optimize low surface brightness observations.

\articlefigure[width=\textwidth]{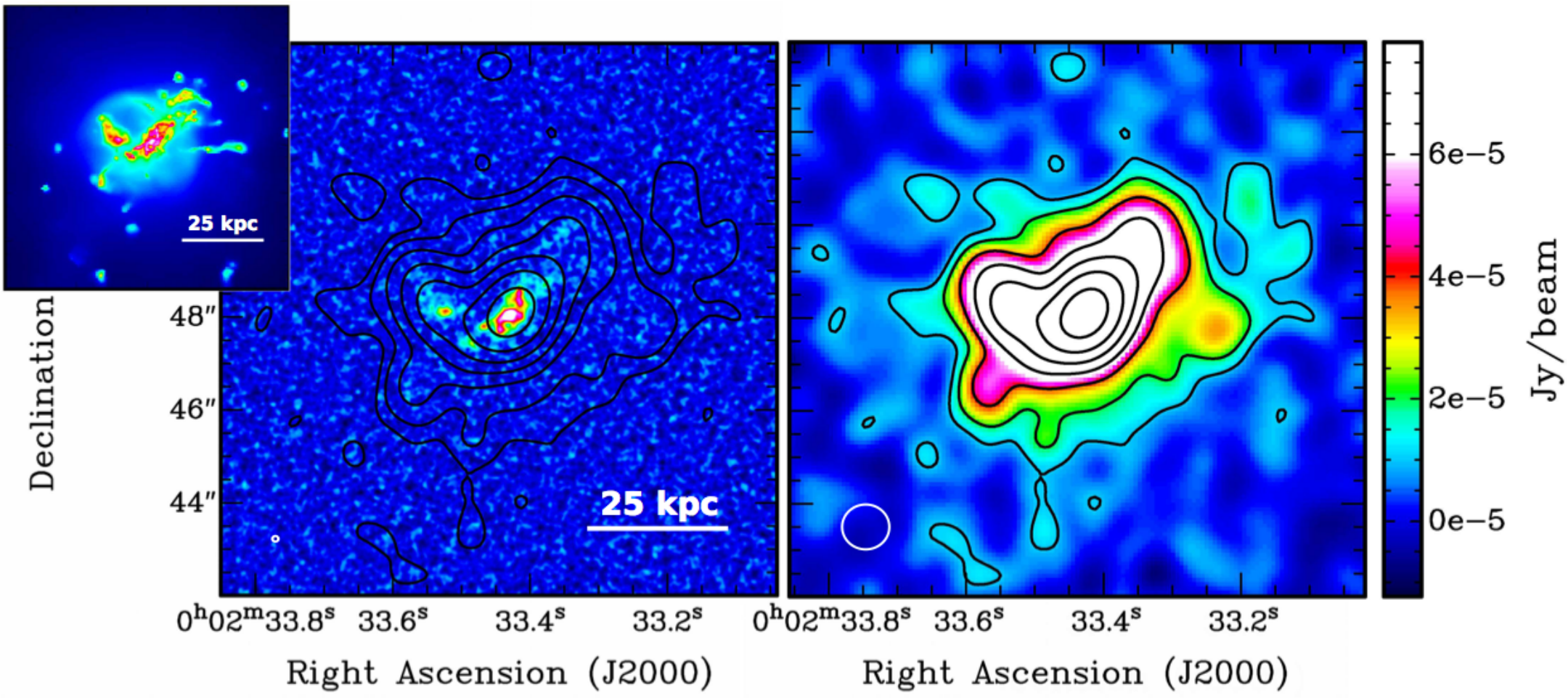}{fig:simulation}
{Simulated CO(1-0) observation of a merging system with extended halo from the cosmological simulations of \citet{nar15} (top-left inset), using the ngVLA reference design of \citet{car18}. The simulations were performed in CASA, and assume $z$\,=\,2, exposure time t$_{\rm exp}$\,=\,48h, channel width of 9 MHz ($\sim$ 70 km\,s$^{-1}$), and total CO(1-0) luminosity $I_{\rm CO(1-0)}$\,=\,0.21 Jy\,beam$^{-1}$ km\,s$^{-1}$ (i.e., similar to that of the Spiderweb Galaxy in Fig.\,2; \citealt{emo16}). Noise has been added to the visibilities, based on a theoretical root-mean-square (rms) noise of 2.3\,$\mu$Jy~beam$^{-1}$ when taking the inner 168 antennas of the array and using natural weighting. {\sl Left:} Simulated CO(1-0) image at a resolution of 0.126$^{\prime\prime}$ ($\sim$1\,kpc), using a robust weighting of 0 and uv-tapering of 0.09$^{\prime\prime}$. This taper provides a resolution relevant only to the antennas that are distributed across the Plains of San Augustin with baselines $\le$30\,km, while down-weighting the core. The synthesized beam is shown in white in the bottom-left corner. The rms noise level is 3.0 $\mu$Jy\,beam$^{-1}$, with a 3$\sigma$ surface brightness limit of 0.5~K, and 3$\sigma$ limit for H$_{2}$ mass clumps of 5.6\,($\alpha_{\rm CO}$/0.8)\,$\times$\,10$^{8}$ M$_{\odot}$. {\sl Right:} Simulated CO(1-0) image at resolution of 1$^{\prime\prime}$, using a robust weighting of 0.5 and uv-tapering of 0.85$^{\prime\prime}$. This taper provides a resolution relevant to the km-scale core of the array, and down-weights the longer spacings. The synthesized beam is shown in white in the bottom-left corner. The rms noise level is 3.4 $\mu$Jy\,beam$^{-1}$, with a 3$\sigma$ surface brightness limit of 9\,mK. The halo gas is detected at 3$\sigma$ across $\sim$80 kpc, and includes $\sim$60$\%$ more CO(1-0) flux than the 0.13$^{\prime\prime}$ plot on the left. The black contours start at 3$\sigma$ (10\,$\mu$Jy) and increase by factors of 2. For visualization purposes, the same contours are also overlaid in the left plot. More details are given in \citet{car18}.}

A more realistic simulation is shown in Fig.\,4. Here we simulate the expected CO(1-0) emission observed with the ngVLA in a massive (M$_*$\,$\sim$\,10$^{12}$\,M$_\odot$) high-$z$ merger system. The CO luminosities are calculated using the \citet{nar12} scaling relations between $X_{\rm CO}$ and $\Sigma_{\rm H2}$, and the \citet{nar14} model relationship between CO excitation ladders and $\Sigma_{\rm SFR}$. For the purposes of modeling the CO flux, the system is placed at $z$\,=\,2 and has its overall flux scaled to the CO(1-0) intensity observed in the Spiderweb Galaxy (Fig.\,2). The system contains bright merging components across $\sim$30\,kpc and a halo stretching twice as far out. We simulate a 48h observation with the ngVLA \citep{car18}, and apply different tapers to increase the surface brightness sensitivity. When going from a moderately tapered 0.13$^{\prime\prime}$ resolution to a heavily tapered 1$^{\prime\prime}$ resolution, the surface brightness sensitivity increases by almost two orders of magnitude, reaching an rms noise level of 3 mK at 1$^{\prime\prime}$ resolution. At this brightness sensitivity, we detect the CO(1-0) across the halo environment on a scale of $\sim$80 kpc. In the simulated 1$^{\prime\prime}$ image we detect $\sim$60 $\%$ more CO(1-0) flux at a level above 3$\sigma$ than in the 0.13$^{\prime\prime}$ image.

Low-surface-brightness observations of molecular gas with the ngVLA core will be complemented by ALMA, provided that ALMA is used in its most compact configurations, because the baseline length {\it in k$\lambda$} increases rapidly at the higher frequencies. ALMA can target not only the high-$J$ transition of CO, but also atomic carbon [C\,{\small I}], which is expected to be concomitant with CO(1-0) and can serve as an important alternative mass tracer \citep[][see also Fig.\,1]{pap04}. With the advent of Band 1 \citep{hua16}, ALMA will be able to observe down to 35 GHz, where it can target CO(1-0) out to $z$\,$\le$2.3. This is also roughly the redshift out to which the Square Kilometre Array (SKA) in its phase-1 will be sensitive for imaging the 21\,cm line of neutral hydrogen in the most massive galaxies. For targeting the low-$J$ CO lines at higher redshifts (15-50 GHz regime), where we can unravel the cold gas content of the very Early Universe, the ngVLA is needed. 

As a note of caution, CO lines become dimmer when observed at higher redshifts as a result of reduced contrast against the increasing temperature of the Cosmic Microwave Background \citep{dac13,zha16}. For extended CO(1-0), this dimming can reach anywhere from the typical 30-40$\%$ expected for the coldest (T$\sim$20\,K) component of the ISM at $z$\,$\sim$\,2 to up to factors of a few for gas in the CGM at higher redshifts. A detailed investigation into the dimming of the low-$J$ CO lines is beyond the scope of this paper. However, to account for this dimming by the CMB, the vastly increased sensitivity of the ngVLA core over existing instruments, like the VLA, is critical. \\
\vspace{-2mm}\\
\noindent{\sl Therefore, the ngVLA will be the only instrument in the foreseeable future than can target the crucial CO(1-0) transition at $z$\,$\ge$\,2.3 with a core-configuration that has the brightness sensitivity to reveal the full extent of the cold molecular Universe.}

\section{Concluding remarks}

While the Next-Generation VLA will be a powerful instrument to study the radio Universe with unprecedented resolution, it is crucial to consider its surface brightness performance during the design phase. The core of the ngVLA, i.e., the baselines in the inner km, will act as a world-leading instrument for unlocking the large-scale cold gas content of the Early Universe. To make optimal use of this capability, it is important to consider the following specifications:

\ \\
\noindent $\bullet$ To maximize the surface-brightness sensitivity of the ngVLA, it is essential to optimize the sensitivity on the shortest ($\le$ several 100m) baselines. 

\ \\
\noindent $\bullet$ To utilize the ngVLA core as a unique stand-alone instrument, it is worth considering sub-array functionality between the inner and outer antennas.

\ \\
Currently, the inclusion of a Short Baseline Array as part of the ngVLA is under consideration. Such a Short Baseline Array aims at recovering emission on the largest spatial scales, and would be analogous to the Atacama Compact Array (a.k.a. Morita Array) of the ALMA observatory. The current reference design envisions that this Short Baseline Array consists of 19 antennas of 6m and 4 total-power dishes of 18m diameter, with interferometric baselines ranging from 11$-$56m \citep[][]{mas18}. It will be essential to investigate to what extent the limited sensitivity of this Short Baseline Array will contribute to efficiently mapping faint, low-surface-brightness CO emission at high redshifts, but this is beyond the scope of this paper.

\ \\
\noindent Concluding, the core of the ngVLA will reveal the cold molecular medium in the Early Universe to limits and at scales not currently detectable with existing millimeter observatories. As such, it will be the most powerful instrument for investigating the cold baryon cycle that drives the early evolution of galaxies and clusters.\\

\acknowledgements We thank Brian Mason for his feedback on the Short Baseline Array. The National Radio Astronomy Observatory is a facility of the National Science Foundation operated under cooperative agreement by Associated Universities, Inc.  



\end{document}